# On the Hotspot Problem in Flash Sintering

Yanhao Dong (dongya@seas.upenn.edu)

*Department of Materials Science and Engineering, University of Pennsylvania,*

*Philadelphia, PA 19104, USA*

**Abstract**

A perturbation analysis has been conducted to evaluate the generation of hotspots inside an electrical loaded ceramic sample, which is assumed to have an Arrhenius-type conductivity. The results identified a critical size, above which a small temperature perturbations will be magnified and hotspots will be generated. It provides an estimate for the largest sample size suitable for flash sintering, beyond which hotspots are likely to form, resulting in inhomogeneous heating and sintering.

## I. Introduction

Field assisted sintering for ceramic materials has attracted much attention in recent years, among which flash sintering emerges since it has been firstly demonstrated by Raj in 2010[1-3]. In a typical flash sintering experiment, a ceramic green body is hold under a constant voltage while the furnace temperature is ramped up. A sudden increase in densification rate and conductivity happens when the furnace reaches a critical temperature. It differs with another popular field assisted sintering technique spark plasma sintering (SPS)[4] from the aspects that (i) flash sintering requires the electric current directly passing through the sample while in SPS electric current by-pass the sample through the highly conducting graphite die and (ii) the former does not involve a high mechanical pressure. As has been explained in the literature[5-10], the fortune to have a flash and a power surge in flash sintering comes from the Arrhenius-type dependence of conductivity on temperature in ceramic materials—Joule heating raises the sample's temperature, which lowers the conductivity, increases the Joule heating power under a constant voltage and thus forms a positive feedback and an unstable heating—and the resultant thermal runaway is likely to be responsible for the rapid densification process. The simplest mathematical treatment for the thermal runaway problem approximated the sample to have an effective temperature, which is a uniform average quantity. The accurate solution also encounters a temperature gradient from the hotter inner part to the colder outer surface, which can be obtained by finite element method[11-13].

In addition to the "global" thermal runaway which rapidly increases the sample's overall temperature, thermal runaway can take place locally and generate hotspots. Such a hotspot problem is well known for microwave heating/sintering[14,15], where sample's permittivity increases with temperature. The dissipated heat would increase the sample's temperature and permittivity, which in turn causes more heat dissipation. Therefore, it shares the essential feature as flash sintering, despite a different description for the power dissipation. (Indeed, flash microwave sintering has also been demonstrated in the literature[16,17].) In the present note, we conducted a simple perturbation analysis on the hotspot problem in flash sintering. The critical size of the hotspot is estimated from experimental conditions. If the sample's dimension is larger than the critical size, hotspots are likely to form, leading to inhomogeneous heating and sintering.

## II. Formulation of the problem

Consider a sample under an applied electric field along $z$-direction and assume the sample is uniform along $z$-direction so that the problem can be simplified to be two-dimensional. The heat equation can be written as

$$\rho c \frac{\partial T}{\partial t} = \frac{E^2}{\rho_0} \exp\left(-\frac{E_a}{k_B T}\right) + \kappa \left(\frac{\partial^2 T}{\partial x^2} + \frac{\partial^2 T}{\partial y^2}\right) \qquad (1)$$

where $\rho$ is the density; $c$ is the heat capacity; $t$ is the time, $T(x, y, t)$ is the temperature profile which varies with $x$, $y$ and $t$; $E$ is the electric field strength; $\rho_0$ and $E_a$ are the pre-exponential and activation energy for sample's resistivity, respectively; $\kappa$ is the thermal conductivity, which for simplicity is assumed to be a constant. Together with the boundary conditions at the sample's surface and initial condition at $t=0$, Eq. (1) gives the transient and steady-state solutions for $T(x, y, t)$. Now assume $T_0(x, y, t)$ is the solution for the uniform heating problem, where $T_0$ decreases monotonically from the inner core of the sample to the outer surface. Hence, it holds for any $x$, $y$ and $t$

$$\rho c \frac{\partial T_0}{\partial t} = \frac{E^2}{\rho_0} \exp\left(-\frac{E_a}{k_B T_0}\right) + \kappa \left(\frac{\partial^2 T_0}{\partial x^2} + \frac{\partial^2 T_0}{\partial y^2}\right) \qquad (2)$$

Now we introduce a small perturbation to $T_0(x, y, t)$ and let

$$T = T_0 + \alpha \sin\left(\frac{x}{\lambda}\right) \sin\left(\frac{y}{\lambda}\right) \exp(-\beta t) \qquad (3)$$

where $0 < \alpha \ll T_0$. Obviously, the perturbation will cease if $\beta > 0$ and magnify if

$\beta < 0$. Therefore, we next seek to determine the sign of $\beta$. Denoting

$$f(\alpha) = \exp\left(-\frac{E_a}{k_B T}\right) = \exp\left(-\frac{E_a}{k_B\left[T_0 + \alpha \sin\left(\frac{x}{\lambda}\right)\sin\left(\frac{y}{\lambda}\right)\exp(-\beta t)\right]}\right) \quad (4)$$

we have

$$f(0) = \exp\left(-\frac{E_a}{k_B T_0}\right) \quad (5)$$

$$f'(\alpha) = \frac{E_a}{k_B T^2}\exp\left(-\frac{E_a}{k_B T}\right)\sin\left(\frac{x}{\lambda}\right)\sin\left(\frac{y}{\lambda}\right)\exp(-\beta t) \quad (6)$$

$$f'(0) = \frac{E_a}{k_B T_0^2}\exp\left(-\frac{E_a}{k_B T_0}\right)\sin\left(\frac{x}{\lambda}\right)\sin\left(\frac{y}{\lambda}\right)\exp(-\beta t) \quad (7)$$

Expanding $\exp\left(-\frac{E_a}{k_B T}\right)$ in Tylor's expansion for $\alpha$ and dropping the higher order terms, we obtain

$$\exp\left(-\frac{E_a}{k_B T}\right) = \exp\left(-\frac{E_a}{k_B T_0}\right) + \alpha \frac{E_a}{k_B T_0^2}\exp\left(-\frac{E_a}{k_B T_0}\right)\sin\left(\frac{x}{\lambda}\right)\sin\left(\frac{y}{\lambda}\right)\exp(-\beta t) \quad (8)$$

Plugging Eq. (3) into Eq. (1) and utilizing Eq. (2) and (8), we obtain

$$\beta = \frac{1}{\rho c}\left[\frac{2\kappa}{\lambda^2} - \frac{E^2}{\rho_0}\exp\left(-\frac{E_a}{k_B T_0}\right)\frac{E_a}{k_B T_0^2}\right] \quad (9)$$

Therefore,

$$\beta < 0, \text{ if } \lambda > \frac{2\kappa}{\frac{E^2}{\rho_0}\exp\left(-\frac{E_a}{k_B T_0}\right)\frac{E_a}{k_B T_0^2}} \quad (10)$$

where the perturbation will increase its magnitude with time and hotspots will be generated.

$$\beta > 0, \text{ if } \lambda < \frac{2\kappa}{\frac{E^2}{\rho_0}\exp\left(-\frac{E_a}{k_B T_0}\right)\frac{E_a}{k_B T_0^2}} \quad (11)$$

where the perturbation will decay with time and a uniform heating is ensured.

## III. Estimation with experimental conditions

Our analysis identified a critical size, above which a small perturbation can bifurcate the solution of a stable heating. In practice, such a perturbation could come from inhomogeneity in either the microstructure (e.g. non-uniform powder packing and sintering) or the electrical loading (e.g. non-uniform contacts between the electrode and the sample). To estimate such a critical size, we used a thermal conductivity of $\kappa=1.7$ W·m$^{-1}$·K$^{-1}$, a Joule heating power density of $\frac{E^2}{\rho_0}\exp\left(-\frac{E_a}{k_B T_0}\right)$ in the range of 100-1,000 mW/mm$^3$, an activation energy $E_a$ of 0.8 eV and $T_0$ around 1,000 °C for YSZ, which gives $\lambda$ of 0.8-1.6 mm. It has two implications. On one hand, it suggests an inherent heating instability for flash sintering, which if applied to large samples non-uniform heating can be triggered. For the above experimental conditions specified for YSZ, the sample size suitable for flash sintering could be limited a few mm. On the other hand, any structural inhomogeneity with a length scale less than $\lambda$ is unlikely to directly come from the effect of a localized heating. Additional effects such as reduction[18] are required to further narrow down the critical size $\lambda$. (In principle, one can modify the expression $\rho_0\exp\left(\frac{E_a}{k_B T}\right)$ for the resistivity to include such addition effects.) However, it should be noted that the above estimation is based on the average Joule heating power, which is defined as the total Joule heating power divided by the total volume. In reality, the temperature distribution is not uniform and the inner core is hotter than the outer surface. Therefore, the inner core has a higher Joule heating power and hence a smaller $\lambda$.

IV. Conclusions

The inherent heating instability in flash sintering has been analyzed by a perturbation method. The results identified a critical size, above which a small temperature perturbations will be magnified and hotspots will be generated. It may limit the largest sample size suitable for flash sintering, beyond which inhomogeneous heating and sintering is likely to happen.